\documentclass[prb,twocolumn,superscriptaddress,showpacs,amsmath,amssymb,amsthm]{revtex4}
\usepackage{amsmath,amssymb,color,graphicx,fancyhdr,hyperref,amsthm,units}

\usepackage{graphicx}
\usepackage{latexsym}
\usepackage{amsmath}
\usepackage{amssymb}
\usepackage{amsfonts}
\usepackage{color}
\usepackage{bm}
\usepackage{verbatim}

\begin{document}

\title{Magnetic behaviour of dirty multiband superconductors near the upper critical field.}

\date{\today}

 \begin{abstract} 
 Magnetic properties of dirty multiband superconductors near the upper critical field are studied. 
 The parameter $\kappa_2$ characterizing magnetization slope is shown 
 to have a significant temperature variation which is quite sensitive to the pairing interactions and relative strengths
 of intraband impurity scattering. In contrast to single-band superconductors the increase of $\kappa_2$ at low temperatures 
 can be arbitrary large determined by the ratio of minimal and maximal diffusion coefficients in different bands. 
 Temperature dependencies of $\kappa_2(T)$ in two-band MgB$_2$ and iron-based superconductors 
 are shown to be much more sensitive to the multiband effects than the upper critical field $H_{c2}(T)$. 
 \end{abstract}

\author{Mikhail~Silaev}
\affiliation{Department of Theoretical Physics and  Center for Quantum Materials,
KTH-Royal Institute of Technology, Stockholm, SE-10691 Sweden}

 \maketitle
 \section{Introduction}

 Recently a number of multiband superconductors have been discovered where the pairing of electrons is supposed to take place 
 simultaneously in several bands overlapping at the Fermi level 
 \cite{MgB2:Discovery,MgB2:MazinPRL,MgB2:Mazin,Sr2RuO4,FeAsExp1,MazinSpm,FeAsMultiband,ChubukovSpm}. 
 One of the first such superconductors found was MgB$_2$ \cite{MgB2:Discovery} which 
 has two distinct superconducting gaps residing on different sheets of the Fermi surface\cite{MgB2:MazinPRL,MgB2:Mazin,MgB2:abinitio}. 
 Up to date MgB$_2$ has the highest critical temperature $T_c = 40$ K among simple binary compounds\cite{MgB2:TopicalReview}.  
 Later on multiband superconductivity has been established in 
 iron-pnictides\cite{FeAsMultiband,ChubukovSpm,MazinSpm}. 
 There a strong interband interaction mediated by antiferromagnetic excitations has been suggested 
 to play the dominant role in pairing resulting in the peculiar $s_\pm$ symmetry of the order parameter. 
 The highest $T_c$ among iron-pnictides is above $100$ K detected in
 atomically thin films of FeSe \cite{FeSeThinFilms}.  
 
 Besides their high $T_c$ both the two-gap MgB$_2$ and iron-based superconductors have 
 remarkable magnetic properties.  The possibility of type-1.5 superconductivity has been suggested 
 to explain vortex clusterization detected in MgB$_4$\cite{Moshchalkov15Main, Moshchalkov15}
 and  Sr$_2$RuO$_4$\cite{Bending15}. The observed unconventional vortex patterns  can result from a 
 non-monotonic vortex interaction generated by the interplay of multiple superconducting coherence lengths 
 in multicomponent superconductors \cite{BabaevSpeight,SilaevBabaevPRB2011,CarlstromPRL2010}.  
  
 In pulsed field experiments the critical fields up to $60$ T have been observed in iron 
 pnictides\cite{GurevichFeAsNature,Exp60T-1,Exp60T-2,Exp60T-3,Exp60T-4,Exp60T-5}. 
 Such values of $H_{c2}$ are high enough to reach a paramagnetic 
 limit\cite{Exp60T-2,Paramagnetic-1,Paramagnetic-2,Paramagnetic-3,Paramagnetic-4,Paramagnetic-5} leading to 
 the possibility of the Fulde-Ferrel-Larkin-Ovchinnikov transition \cite{GurevichFFLO}.  
 Comparably large upper critical fields were obtained by introducing disorder in 
 MgB$_2$ thin films where the record values of $H^\perp_{c2}=35$T and $H_{c2}^\parallel=51$T
 were observed perpendicular and parallel to the crystal anisotropy $ab$ plane respectively\cite{MgB2Ex1}.    
  
 The interplay of several pairing channels in multiband superconductors was predicted to produce convex-shaped
 temperature dependencies of the upper critical field \cite{GurevichMgB2,GurevichPhysicaC,KoshelevGolubovHc2,GurevichFFLO}.
 Due to their anomalous shapes the $H_{c2}(T)$ curves reach much larger values at $T\to 0$  
 than was expected from a one-gap theory  $H_{c2}(0)=0.69 T_c H_{c2}(T_c)$
 \cite{SingleBandHc2Maki,SingleBandHc2Werthamer}. This explains an enormous enhancement of the 
 upper critical field in MgB$_2$ by non-magnetic impurities\cite{MgB2Ex2,MgB2Ex3,MgB2Ex4}. 
 Experiments measuring upper critical fields in  the disordered  MgB$_2$\cite{MgB2Ex1} and 
 certain iron-based superconductors \cite{GurevichFeAsNature} are consistent with  theoretical 
 calculations using a two-gap model \cite{GurevichMgB2,GurevichPhysicaC,GurevichFFLO}.
 Therefore a convex shape of $H_{c2}(T)$  dependence is considered as one of the hallmarks of multiband 
 pairing\cite{GurevichFeAsNature,Hc2SrEu,Hc2CaLaFeAs,BalatskyEdge,Paramagnetic-3}  . 
 However it is not a universal feature of multiband superconductors since concave $H_{c2}(T)$
 curves were observed in MgB$_2$ without artificially introduced disorder\cite{MgB2:TopicalReview} 
 as well as in many iron-pnictide compounds\cite{Exp60T-3,Exp60T-4,Paramagnetic-4,concave-1}.
  
 In order to find a robust test for mutiband pairing it is natural to look for 
 an unusual behaviour of magnetization at $H<H_{c2}$. In the vicinity of $H_{c2}$ magnetization $M_z$ 
 can be characterized by the parameter $\kappa_2(T)$ 
 introduced by Maki \cite{Maki}
 \begin{equation}\label{Eq:Magnetization}
 M_z = \frac{H-H_{c2}}{4\pi\beta_L(2\kappa^2_{2}-1)},
 \end{equation} 
 where $z$-axis is directed along the magnetic field,
 $\beta_L$ is an Abrikosov parameter equal to $1.16$ for a triangular lattice\cite{AbrikosovParameter}.
  In single-band superconductors the parameter $\kappa_2$ has been studied extensively in the 
 clean \cite{MakiTsuzukiKappa2} and dirty limits\cite{Caroli,UsadelKappa2}, 
 for the arbitrary strength of impurity scattering\cite{EilenbergerKappa2,KitaKappa2} and 
 taking into account strong electron-phonon coupling effects\cite{UsadelKappa2Strong}.
 Dirty single-band superconductors were shown to have a universal behaviour characterized by 
 a slow monotonic increase of $\kappa_2$  \cite{Maki, Caroli,UsadelKappa2} with cooling from 
 $\kappa_{2}(T=T_c)=\kappa_{GL}$  to $\kappa_{2}(T=0)\approx 1.2\kappa_{GL}$,
 where $\kappa_{GL}$ is the Ginzburg-Landau parameter at $T=T_c$.
 The theoretical calculations  were found to be in good agreement with experimentally measured 
 $\kappa_2(T)$ dependencies in several superconducting 
 alloys.\cite{Kappa2Exp1,Kappa2Exp2,Kappa2Exp3} 
 
 The parameter $\kappa_2$ is a basic quantity of type-II superconductors determining their thermodynamic \cite{MakiSpecificHeat}
 and transport properties\cite{CaroliMakiFluxFlow,ThompsonFluxFlow} near $H_{c2}$.
 However the theory calculating $\kappa_{2}$ in multiband superconductors has been lacking.  In the present paper
 we demonstrate that this parameter is much more sensitive to multiband effects than the upper critical field. 
 In MgB$_2$ and iron pnictides $\kappa_2(T)$ dependencies are shown to reveal pronounced signatures of multiband pairing in the regimes when 
 $H_{c2}(T)$ curves deviate only slightly from the conventional single-band behaviour. 
 The $\kappa_2 (T)$ anomalies signal unconventional thermodynamic and transport characteristics of multiband superconductors.
   
 The structure of this paper is as follows. In Sec.(\ref{Sec:MultibandUsadel}) basic equations of the multiband Usadel theory 
 are introduced. General formulas describing the high-filed magnetic response of dirty multiband superconductors are derived in 
 Sec.(\ref{Sec:Results}) including equations for the $H_{c2}$ in Sec.(\ref{Subsec:ResultsHc2})
 and the magnetization in Sec.(\ref{Subsec:ResultsMagnetization}). Several examples of two-band superconductors are 
 considered in Sec.(\ref{Sec:Examples}). Results are discusses in Sec.(\ref{Sec:Discussion}) and conclusions are given in Sec.(\ref{Sec:Conclusion}).        
      
 \section{Multiband Usadel theory.} \label{Sec:MultibandUsadel}
    
 We consider multiband superconductors in a dirty limit using the Usadel theory\cite{GurevichMgB2}. Each $k$-th band 
 is described in terms of the quasiclassical Green's function matrix $\hat g_k= \hat g_k (\varepsilon, {\bm r})$ which is defined 
 as follows
 \begin{equation}\label{Eq:GFNambu}
 \hat g_k = \left( g_k \;\;\;\;\;\; f_k \atop -f^+_k \; -g_k \right)
 \end{equation}
 and subject to the normalization constraint $\hat g_k^2=1$. The matrix Usadel equation reads\cite{Schon,GurevichMgB2}
 \begin{equation}\label{Eq:UsadelGen}
  D_k \hat\partial_{\bm r} (\hat g_k\hat\partial_{\bm r}\hat g_k) - [\omega\tau_3 + \hat\Delta_k, \hat g_k] =0 
 \end{equation}
  where $D_k$ is the diffusion constant, and  
 $\hat\Delta_k ({\bm r})= |\Delta_k| \tau_2 e^{-i\theta_k \tau_3}$ 
 is the matrix gap function in $k$-th band. 
 In Eq.(\ref{Eq:UsadelGen}) the covariant differential superoperator is defined by 
 $\hat \partial_{\bf r} \hat g= \nabla \hat g -ie{\bm A} \left[\tau_3, \hat g \right] $.
 The 12 component of the matrix Eq.(\ref{Eq:UsadelGen}) yields:    
   \begin{equation} \label{Eq:Usadel12}
   \frac{D_k}{2i} ( g_k \bm {\hat \Pi}^2 f_k  - f_k \nabla^2 g_k ) = \Delta_k g_k -i\omega f_k 
   \end{equation}
 where $\bm {\hat \Pi} = \nabla - 2ie {\bm A}$. Similar equation given by the 21 component of (\ref{Eq:UsadelGen}) 
 yields $f^+({\bm r},\omega)=-f^*({\bm r},\omega)$. 
  The gap in each band is determined by self-consistency equations
 \begin{equation}\label{Eq:SelfConst}
 \Delta_k (\bm r) = 2i\pi T  \sum_{j=1}^N\sum_{n=0}^{N_D} \lambda_{kj}  f_j (\omega_n) 
 \end{equation} 
  where $\hat\lambda $ is the $N\times N$ coupling matrix satisfying general symmetry relations $\nu_k\lambda_{kj}=\nu_j\lambda_{jk}$ and the sum by Matsubara frequencies $\omega_n=(2n+1)\pi T$ is taken in the limits $N_D(T)=\Omega_D/(2\pi T)$ set by the Debye frequency $\Omega_D$.     
 The electric current density is given by 
 \begin{equation}\label{Eq:Current}
 {\bm j}  =  i\pi T \sum_{k=1}^N\sum_{n=0}^{\infty}  
 \frac{\sigma_k}{e} {\rm Tr} [\tau_3 \hat g_k(\omega_n) \hat\partial_{\bm r} \hat g_k (\omega_n) ]
 \end{equation} 
 where the partial conductivities are $\sigma_k=2e^2\nu_k D_k $ and $\nu_k$ are the densities of states per one spin projection.
 The sum over frequencies in Eq.(\ref{Eq:Current}) converges therefore no cutoff is needed.   
 The magnetization of a superconducting sample ${\bm M}$ is determined by the current (\ref{Eq:Current}) according to the usual 
 relation  $\nabla \times {\bm M} = {\bm j}$.
                           
 \section{Multiband superconductors in large magnetic fields. } \label{Sec:Results}  
 \subsection{The upper critical field $H_{c2}$.} \label{Subsec:ResultsHc2}
 At large magnetic fields $H_{c2}- H \ll H_{c2}$ we can apply approximations related to the smallness of the order parameter  
 $|\Delta_k|\propto \sqrt{ 1- H/H_{c2}}$. To calculate the structure of a vortex lattice in a two-band superconductor let us consider the 
 linear integral-differential system consisting of Usadel Eqs.(\ref{Eq:Usadel12}) linearized with respect to the normal state solution
  \begin{equation} \label{Eq:UsadelLinearized} 
 \hat L_\omega f_k = i\Delta_k  ; \;\;  \hat L_\omega = \frac{D_k}{2}{\bm {\hat \Pi}}^2_0 - |\omega|,
 \end{equation}  
 supplemented by the self-consistency relation (\ref{Eq:SelfConst}). 
 In a linearised theory the magnetic filed is not perturbed by the vortex currents therefore we put 
 ${\bm B}_0= H_{c2} {\bm z}$ and choose a Landau gauge in Eq.(\ref{Eq:UsadelLinearized}) 
 ${\bm A}_0=  H_{c2} x{\bm y}$. Then the gradient term in Eq.(\ref{Eq:UsadelLinearized}) is $\bm {\hat \Pi}_0 = \nabla - 2ie {\bm A}_0$.
 A periodic vortex lattice is described by the Abrikosov 
 solution of Eqs.(\ref{Eq:UsadelLinearized},\ref{Eq:SelfConst}) which in general has the following form 
 \begin{align}\label{Eq:Ansatz}
 &\Delta_k (\bm r) = \Delta b_{k} \Psi(\bm r)   \\
 &\Psi(\bm r) = \sum_n C_n e^{ i n py}  \Psi_0 (x-nx_0) \label{Eq:Abrikosov}
 \end{align} 
 where $|C_n|=1$, $x_0= p/(2eH_{c2})$ and parameter $p$ is determined by the lattice geometry.
 The lowest Landau level wave function 
 $\Psi_0(x) =  2L_H\sqrt{\pi}\exp(-x^2/2L_H^2)$ satisfies 
 $(L_H^{2}\partial_x^2 - x^2/ L_H^{2} + 1 )\Psi_0 = 0 $
 where the magnetic length is $L_H=1/\sqrt{2eH_{c2}}$. The gaps $\Delta_k$ are determined by the common amplitude $\Delta$
 and a normalized set of components $\sum_k b_k^2 =1$. 
 
 The solution of Eq. (\ref{Eq:UsadelLinearized}) yields 
 \begin{equation} \label{Eq:Coeffitients}
 f_k (\bm r,\omega_n) = \frac{\Delta_k (\bm r)}{i(q_k+|\omega_n|)}  
 \end{equation}
 where $q_k = e H_{c2} D_k$. 
 Substituting the ansatz (\ref{Eq:Ansatz},\ref{Eq:Coeffitients}) 
 to the self-consistency Eq.(\ref{Eq:SelfConst}) we get the homogeneous linear system 
 $\hat A (b_1,.. b_N)^T =0$ for the order parameter amplitudes where  
 \begin{equation} 
 \label{Eq:Hc2}
 \hat A = \hat \Lambda^{-1} -  \hat I \left[ G_0+ \ln (T_c/T ) + \psi (1/2) - \psi ( 1/2+\hat\rho ) \right].  
 \end{equation}
 Here $\psi(x)$ is a di-gamma function and the diagonal matrix $\hat\rho$ is given by  $(\hat \rho )_{ij} = \delta_{ij} q_i/(2\pi T)$.
 The solvability condition ${\rm det}\hat A =0$
 determines the upper critical field $H_{c2}$ of a dirty multiband superconductor.
     
   It is instructive to consider in more detail Eq.(\ref{Eq:Hc2}) in two-band superconductors. In this case 
   $G_0=  ({\rm Tr}\;\Lambda - \lambda_0 )/w$ where 
   $w=\lambda_{11}\lambda_{22} - \lambda_{12}\lambda_{21}$, $\lambda_0 =\sqrt{ \lambda_-^2 +4\lambda_{12}\lambda_{21}}$
   and $\lambda_- = \lambda_{11} - \lambda_{22}$.
   The equation ${\rm det} \hat A =0$ can be resolved in terms of the 
   $\ln (T/T_c)$ yielding in general two different solutions
   \begin{align} \label{Eq:Hc2PositiveW}
   \ln (T/T_c) = &- (U_1+U_2 + \lambda_0/w)/2  \\ \nonumber
   &+ [ (U_1-U_2-\lambda_-/w)^2/4 + \lambda_{12}\lambda_{21}/w^2 ]^{1/2} \\ \label{Eq:Hc2NegativeW} 
   \ln (T/T_c) = &- (U_1+U_2 + \lambda_0/w)/2  \\ \nonumber
   &- [ (U_1-U_2-\lambda_-/w)^2/4 + \lambda_{12}\lambda_{21}/w^2 ]^{1/2} ,
   \end{align}       
 where $U_{k}= \psi( 1/2+\rho_k) - \psi(1/2)$. 
 Taking the limit $T\to T_c$ one can see that the physical 
 solutions are {\bf (i)} (\ref{Eq:Hc2PositiveW}) in case when $w \equiv {\rm det} \hat \Lambda >0$ 
 and {\bf(ii)} (\ref{Eq:Hc2NegativeW}) in case when $ w\equiv {\rm det} \hat \Lambda <0$ . 
 While the case  {\bf (i)} corresponds to the coupling parameters of MgB$_2$\cite{GurevichMgB2,GurevichPhysicaC} the case    
 {\bf(ii)} describes multiband superconductors with interband-dominated pairing when 
 $\lambda_{12}\lambda_{21}> \lambda_{11}\lambda_{22}$ such as iron-pnictide compounds\cite{ChubukovSpm,MazinSpm,GurevichFeAsNature}.         
   
   \subsection{The magnetization slope $dM_z/dH$.} \label{Subsec:ResultsMagnetization}
   Magnetic field created by vortex currents (\ref{Eq:Current}) can be found using the solution (\ref{Eq:Ansatz}). 
   Taking into account that Green's functions $f_k({\bm r})$ given by (\ref{Eq:Abrikosov}) satisfy the relation
   \begin{equation}\label{Eq:UsefulRelation}
   i\partial_x f_k = (\partial_y -2ie A_y) f_k
   \end{equation}
    we obtain the multi-band expression for magnetization 
   \begin{equation}\label{Eq:MagnetizationMultiband}
   4\pi M_z(\bm r) = -\sum_k \frac{\sigma_k}{eT} \psi^\prime_k|\Delta_k|^2.  
   \end{equation}    
           
  The order parameter amplitudes $\Delta_k$ 
   can be found according to the following straightforward algorithm. 
   First, non-linear corrections $\tilde{f}_k$ are obtained from
   Eq.(\ref{Eq:Usadel12}) taking into account higher-order terms in $\Delta_k$: 
  \begin{align} \label{Eq:fCorrections} 
  & \hat L_\omega\tilde{f}_k= - \frac{i\Delta_k |\Delta_k|^2}{2(q_k+|\omega|)^2} + \\ \nonumber
  & \frac{eD_k \{\bm{\hat \Pi}_0, \bm{A}_1 \} \Delta_k }{(q_k+|\omega|)}- 
  \frac{i( 2q_k\Delta_k |\Delta_k|^2 + D_k \Delta_k \nabla^2|\Delta_k|^2 )}{4(q_k+|\omega|)^3 .} 
  \end{align}
   Then the self-consistency Eq.(\ref{Eq:SelfConst}) yields a non-homogeneous linear system for the corrections $\tilde{\Delta}_j$ :
   \begin{align} \label{Eq:Non-hom}
   \hat C_{kj} \tilde{\Delta}_j =
    2\pi i T \sum_{n=0}^{\infty} \nu_k\tilde{f}_k(\omega_n),  \\ \label{Eq:Operator}
    \hat C=  \hat\nu\hat \Lambda^{-1} - 2\pi i T \sum_{n=0}^{N_D} \hat \nu \hat L_{\omega_n}^{-1} , 
   \end{align}    
   where $\hat\nu_{kj}= \nu_k\delta_{kj}$. Since the matrix $\hat\nu \Lambda^{-1}$ is symmetric the operator $\hat C$
   is hermitian and the linear solution (\ref{Eq:Ansatz})
   belongs to its kernel $\hat C_{kj}\Delta_j =0$. Hence multiplying the l.h.s. of a 
   non-homogeneous Eq.(\ref{Eq:Non-hom}) by $\Delta^*_k$ we get $\sum_{k,j}\langle \Delta^*_k \hat C_{kj}\tilde{\Delta}_j\rangle =0 $. 
   Thus Eq.(\ref{Eq:Non-hom}) is solvable if its r.h.s. is orthogonal to the linear solution
  \begin{equation} \label{Eq:Orthogonality}
  \sum_{k} \sum_{n\geq 0}  \nu_k\langle \Delta_k^* \tilde{f}_k (\omega_n)\rangle  =0.
  \end{equation}
   
  To calculate each term in the sum (\ref{Eq:Orthogonality}) we multiply Eq.(\ref{Eq:fCorrections}) by $\Delta_j^*$ and 
   average over space  coordinates taking into account the relations  
  \begin{align} \label{Eq:Trivial1}
  &\langle \Delta^*_k \hat L_{\omega} \tilde{f}_k \rangle =
  -(|\omega|+q_k) \langle\Delta_k^* \tilde{f}_k \rangle \\ \label{Eq:Trivial2}
  &\langle \Delta_k^*\{\bm {\hat \Pi}_0, {\bm A}_1 \} \Delta_k \rangle = -i \langle B_1|\Delta_k|^2 \rangle \\ \label{Eq:Nontrivial}
  & 2q_k \langle |\Delta_k|^4\rangle + D_k \langle|\Delta_k|^2 \nabla^2|\Delta_k|^2 \rangle =0
  \end{align}
  where $B_1= -\delta H + 4\pi M_z$ and  $\delta H = H_{c2}-H$. 
  The relations (\ref{Eq:Trivial1},\ref{Eq:Trivial2}) can be obtained by a straightforward calculation 
  while (\ref{Eq:Nontrivial})  
  is less trivial although it has been used in the theory of single-band superconductors\cite{BrandtOvchinnikov}.
  The detailed derivation of Eq.(\ref{Eq:Nontrivial}) is shown in the 
  Appendix (\ref{Eq:Appendix}).  
  
  To simplify the further derivation let us consider from the beginning a high-$\kappa$ limit 
  when $\sigma_kD_k\ll 1$. In this case we can neglect the magnetization in Eq.(\ref{Eq:Nontrivial}) 
   to get finally 
  \begin{align} \nonumber
  & 2i\pi^2 T^3 \sum_{n\geq 0} \langle\Delta_k^* \tilde{f}_k(\omega_n) \rangle  = 
   e T D_k\psi^{\prime}_k \delta H \langle |\Delta_k|^2 \rangle - \\ \label{Eq:LinUsadelSolution}
  & 2\sigma_kD_k \psi^{\prime 2}_k \tilde{\kappa}_{k}^2 \langle |\Delta_k|^4 \rangle,
  \end{align}  
   where $\tilde{\kappa}_{k}$ are single-band parameters given by \cite{Maki,Caroli}     
  \begin{equation}\label{Eq:kappaSingleBand}
  \tilde{\kappa}_{k} = \left[ \frac{-\psi^{\prime\prime}_k}{16\pi \sigma_kD_k\psi^{\prime 2}_k} \right]^{1/2}.
  \end{equation}
 Combining Eqs.(\ref{Eq:Orthogonality}) and (\ref{Eq:LinUsadelSolution}) we obtain the order parameter amplitude in Eq.(\ref{Eq:Ansatz})
 given by
 \begin{equation} \label{Eq:OrderParameterAmplitude}
 \Delta= \left[ \frac{eT\delta H}{2\beta_L} \frac{\sum_{k} \nu_k b^2_k D_k\psi^{\prime}_k }
 {\sum_{k} \nu_k b^4_k \sigma_kD_k\psi^{\prime 2}_k \tilde{\kappa}_k^2 }  \right]^{1/2},
 \end{equation}
 where the Abrikosov parameter is $\beta_L = \langle |\Psi|^4\rangle /\langle |\Psi|^2\rangle^2 $. 
  
  The derived amplitude $\Delta$ is a basic parameter for calculations 
 of thermodynamic and transport properties of superconductors near $H_{c2}$. In particular using Eq.(\ref{Eq:MagnetizationMultiband}) 
 we obtain an expression for the space- averaged magnetization $M_z=-\delta H (d M_z/d H)$ where the slope is given by  
 \begin{equation} \label{Eq:MzMultiband}
  \frac{dM_z}{dH} = \frac{ \left(\sum_{k} \sigma_{k} b^2_k \psi^{\prime}_k \right) 
 \left( \sum_{k} \nu_k b^2_k D_k\psi^{\prime}_k \right) }
 {8\pi\beta_L \sum_{k} \nu_k b^4_k \sigma_kD_k\psi^{\prime 2}_k \tilde{\kappa}_k^2 }  .
 \end{equation}
 Comparing Eq.(\ref{Eq:MzMultiband}) with the conventional parametrization (\ref{Eq:Magnetization}) in the limit $\kappa_2\gg 1$ we
 find an effective parameter 
 \begin{equation} \label{Eq:kappa2Res}
 \kappa_2 = \sqrt{ \frac{\sum_{k} \nu_k b^4_k \sigma_kD_k\psi^{\prime 2}_k \tilde{\kappa}_k^2 }
  { \left(\sum_{k} \sigma_{k} b^2_k \psi^{\prime}_k \right) \left( \sum_{k} \nu_k b^2_k D_k\psi^{\prime}_k \right) }} . 
\end{equation}  
 Close to the critical temperature $\kappa_2$ reduces to the Ginzburg-Landau parameter 
 $\kappa_2(T_c)=\kappa_{GL}\equiv \lambda_L/\xi $ where $\lambda_L$ is the London penetration length and  $\xi = 1/\sqrt{2eH_{c2}}$
 is the coherence length. Note that this definition of coherence length is applicable only for dense vortex lattices near $H_{c2}$. 
 The physics of dilute vortex configurations in multicomponent systems is regulated by the interplay of multiple coherence lengths
 resulting in non-monotonic vortex interactions\cite{BabaevSpeight,SilaevBabaevPRB2011,CarlstromPRL2010}. 
  
 \section{Examples of two-band superconductors.}   
  \label{Sec:Examples}

\begin{figure}[h!]
 \centerline{$
 \begin{array}{c}
 \includegraphics[width=0.50\linewidth]{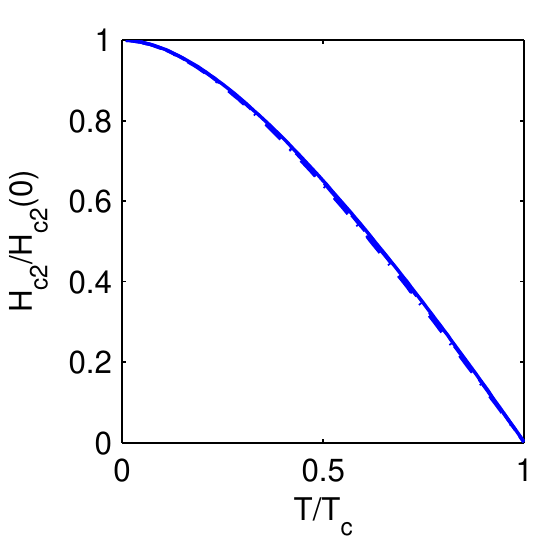}  
 \includegraphics[width=0.50\linewidth]{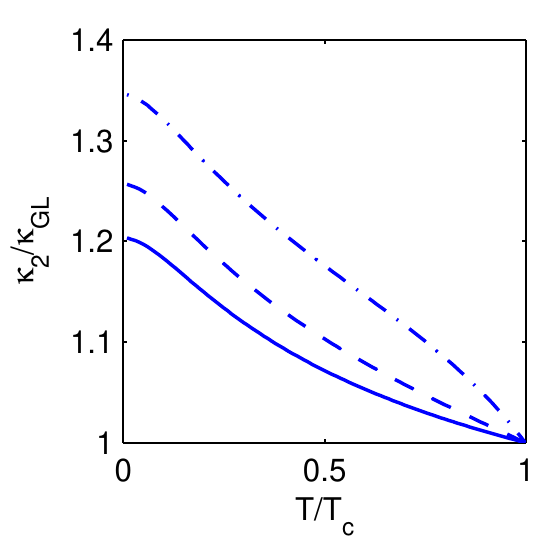}
 \put (-145,100) {(a)}
 \put (-20,100) {(b) }
 \put (-140,123) {$D_1/D_2=1;\; 0.5;\; 0.25$} \\
 \includegraphics[width=0.50\linewidth]{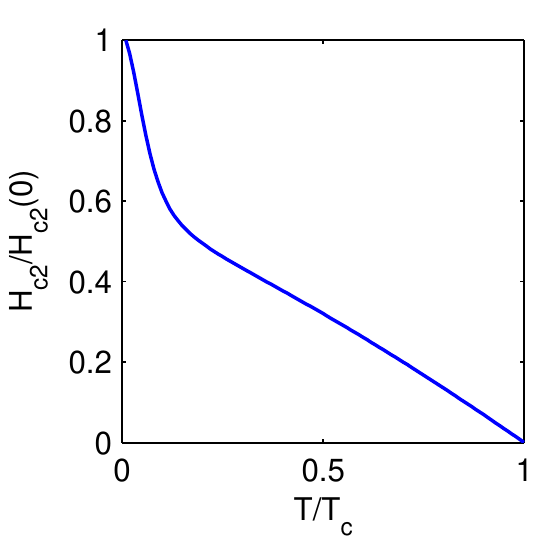} 
 \includegraphics[width=0.50\linewidth]{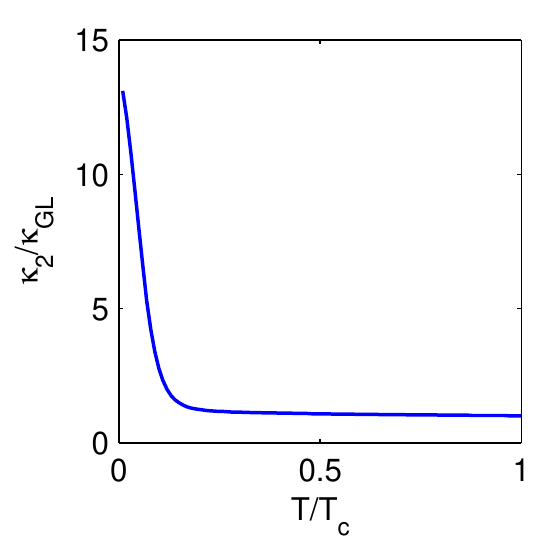} 
 \put (-145,100) {(c)}
 \put (-20,100) {(d)}
 \put (-140,123) {$D_1/D_2=20$} \\
 \includegraphics[width=0.50\linewidth]{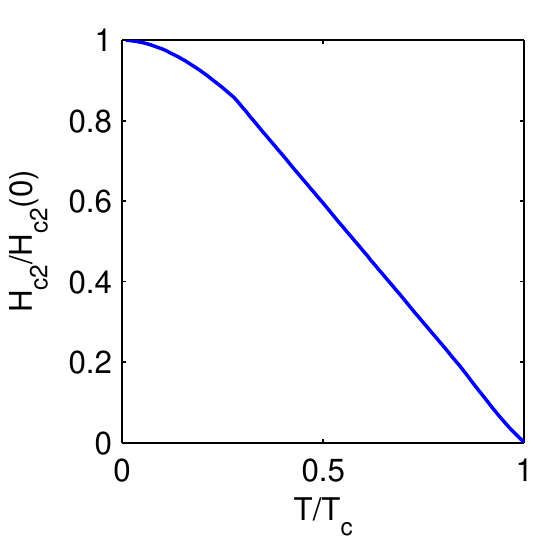} 
 \includegraphics[width=0.50\linewidth]{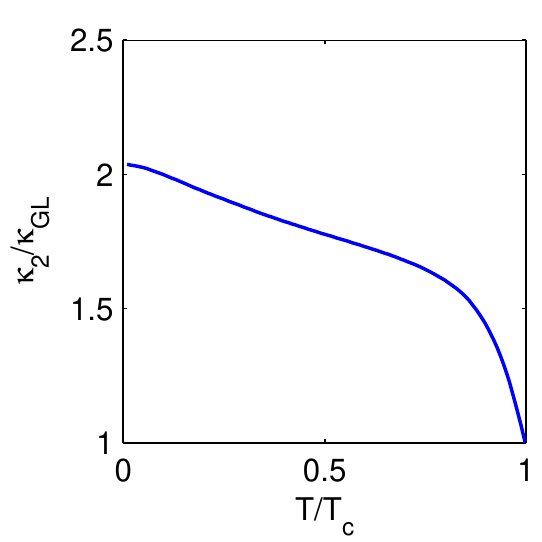}  
 \put (-145,100) {(e)}
 \put (-20,100) {(f) }
 \put (-140,123) {$D_1/D_2=0.05$}  
 \end{array}$}
 \caption{\label{Fig:MgB2} (Color online) Magnetic properties  of the two-band superconductor MgB$_2$ with coupling parameters 
mentioned in the text. 
 The panels show (a,c,e) $H_{c2}(T)$ and (b,d,f) $\kappa_2(T)$ as given by Eqs.(\ref{Eq:Hc2PositiveW},\ref{Eq:kappa2Res}) for different values of the ratio $D_1/D_2$. In (a,b) solid, dashed and dash-dotted lines correspond to $D_1/D_2=1;\;0.5;\;0.25$ respectively. In (a) these curves are almost undistinguishable.  (c,d) $D_1/D_2=20$ and (e,f) $D_1/D_2=0.05$.
   }
 \end{figure}
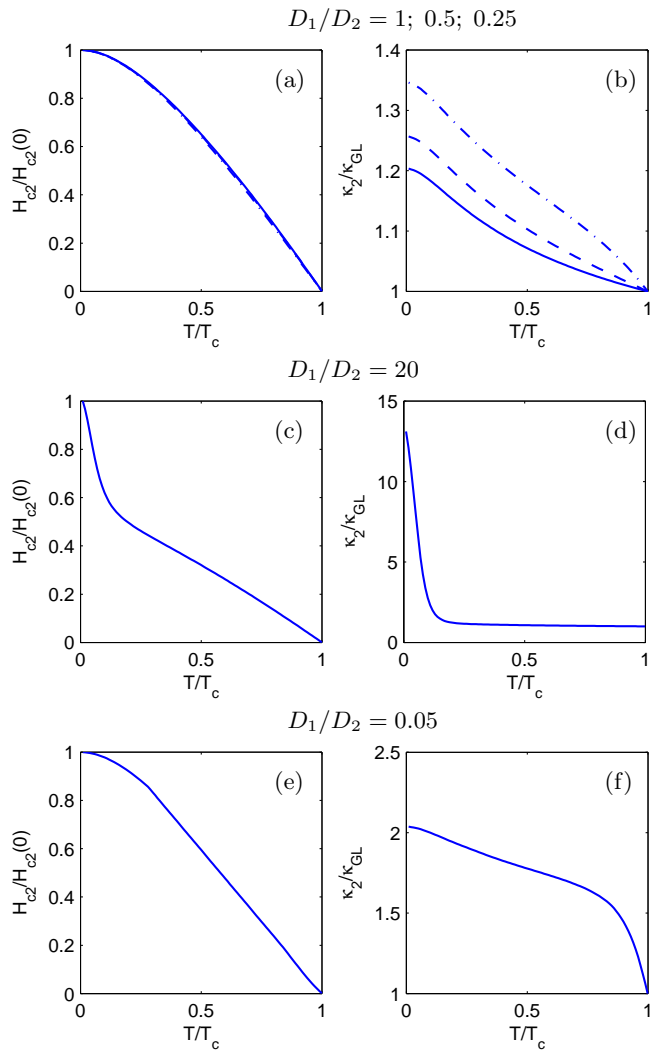

 The general formalism developed in previous sections can be applied to study magnetic properties of particular multiband compounds.
 To begin with we consider a two-band model of MgB$_2$ characterized by the coupling parameters \cite{MgB2:constants}
  $\lambda_{11}=0.81$, $\lambda_{22}=0.285$, $\lambda_{12}=0.119$, $\lambda_{21}=0.09$. 
  Temperature dependencies of $H_{c2}(T)$ and $\kappa_2(T)$ are shown in Fig.(\ref{Fig:MgB2}) for different values of 
  (a) $D_1/D_2=1;\;0.5;\; 0.25$, (b)  $D_1/D_2=20$, (c)  $D_1/D_2= 0.05$.
 For equal diffusion coefficients $D_1/D_2=1$ the single-band
 behaviours\cite{SingleBandHc2Werthamer,SingleBandHc2Maki,Caroli} of $H_{c2}$ and $\kappa_2$ are recovered, see Fig.(\ref{Fig:MgB2})a,b. 
 As will be shown below this result is valid for any number of bands and arbitrary pairing matrix. 
 
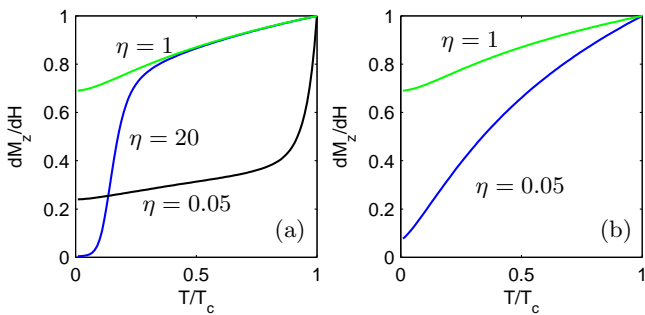
\begin{figure}[htb!]
 \centerline{$
 \begin{array}{c}
 \includegraphics[width=0.50\linewidth]{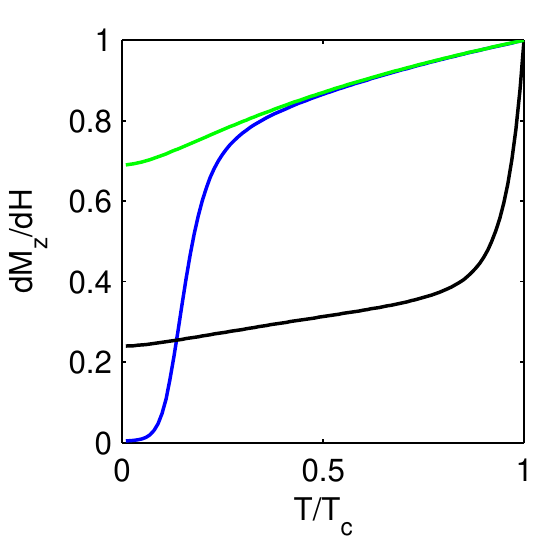}
 \put (-20,30) {(a)}
 \put (-80,100) {$\eta=1$}
 \put (-75,65) {$\eta=20$}
 \put (-70,40) {$\eta=0.05$}  
 \includegraphics[width=0.50\linewidth]{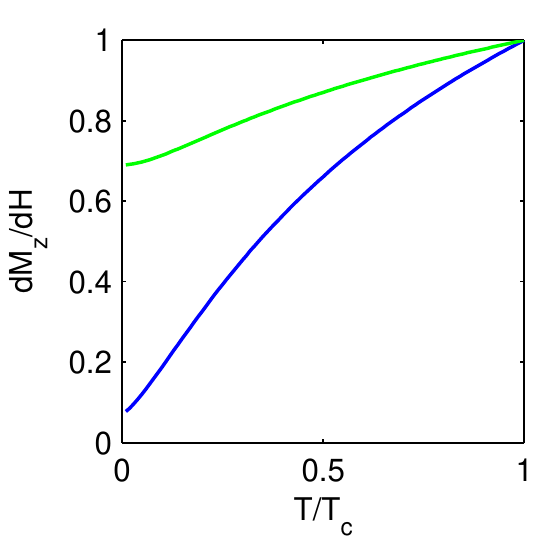}
 \put (-20,30) {(b)}
 \put (-80,102) {$\eta=1$}
 \put (-67,47) {$\eta=0.05$ } 
 \end{array}$}
 \caption{\label{Fig:Magnetization} (Color online) Slopes of the magnetization curves $dM_z/dH$ 
 in two-band superconductors with coupling parameters corresponding to (a) MgB$_2$
  $\lambda_{11}=0.81$, $\lambda_{22}=0.285$, $\lambda_{12} =0.119$, $\lambda_{21}=0.09$ and
 (b) iron-pnictides $\lambda_{11}=\lambda_{22}=0$, $\lambda_{12}=\lambda_{21}=-0.5$. 
 The value of $\eta=D_1/D_2$ is marked near each curve. The slopes are normalized to their values at $T_c$.  }
 \end{figure}

 The disparity of diffusion coefficients $D_1/D_2\neq 1$ 
  results in significant variations of $\kappa_2$.  
   Comparing Figs.(\ref{Fig:MgB2})a and (\ref{Fig:MgB2})b one can see that  $\kappa_2(T)$ is much 
 more sensitive to the ratio of duffusivities than the second critical field. 
  The curvature variations of $H_{c2}(T)$ are noticeable only in the limit $D_2\ll D_1$ as shown 
 Fig.(\ref{Fig:MgB2}c). As demonstrated in the Fig.(\ref{Fig:MgB2}e) the opposite limit $D_1\ll D_2$ yields ordinary concave curves $H_{c2}(T)$  almost within the entire temperature  
 domain except of the small vicinity of $T_c$. 
 On the contrary temperature dependencies of $\kappa_2(T)$ shown for the same parameters in Fig.(\ref{Fig:MgB2}f)
 are drastically different from the single-band case. 
   Of particular interest is a sharp increase of $\kappa_2(T\to 0)$ which is most pronounced under the condition
  $D_2\ll D_1$ relevant to MgB$_2$ \cite{GurevichMgB2} [see Fig.(\ref{Fig:MgB2}d)].
 Physically thus means that the slope of magnetization curve becomes much less 
 steep as shown in the Fig.(\ref{Fig:Magnetization}). 
 The low-temperature increase of $\kappa_2$  can be considered as a feasible probe of 
 the multiband pairing. 
     
 \begin{figure}[htb!]
 \centerline{$
 \begin{array}{c}
 \includegraphics[width=0.50\linewidth]{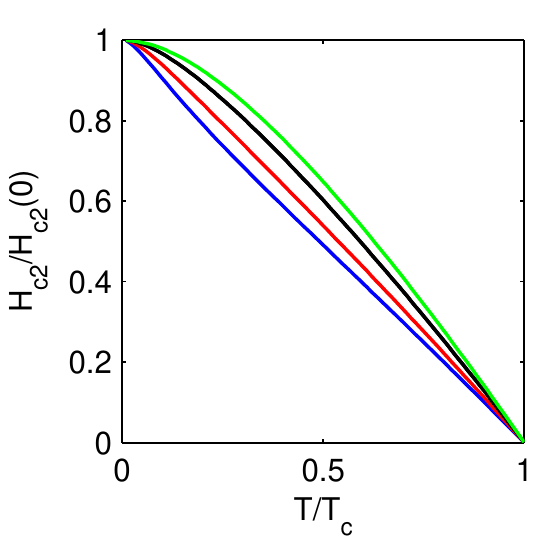}  
 \includegraphics[width=0.50\linewidth]{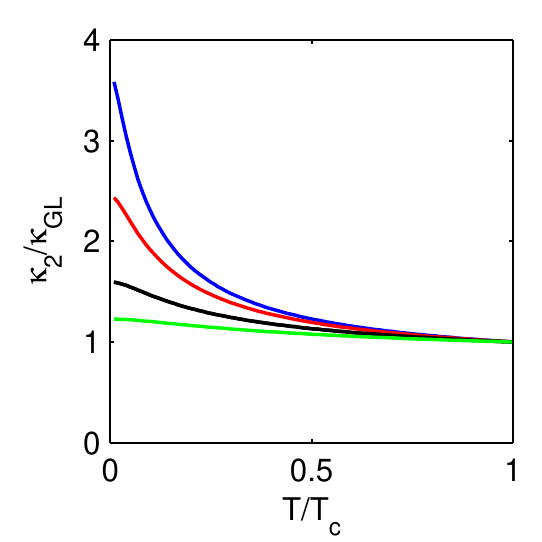}
 \put (-145,100) {(a)}
 \put (-20,100) {(b) }
 \end{array}$}
 \caption{\label{Fig:Spm} (Color online) Magnetic properties of a two-band superconductor with interband- dominated pairing 
 $\lambda_{11}=\lambda_{22}=0$, $\lambda_{12}=\lambda_{21}=-0.5$ corresponding to iron-pnictide superconductors. 
 (a) $H_{c2}(T)$ curves from {\it top to bottom} 
 correspond to $D_1/D_2=1;\;0.25;\;0.1;\; 0.05$. (b) The magnetization parameter 
 $\kappa_2(T)$ as given by the Eq.(\ref{Eq:kappa2Res}). The curves from {\it bottom to top} correspond to the same sequence of
 $D_1/D_2$ as in (a). }
 \end{figure}   

 Next we consider two-band superconductors with pairing from interband repulsion 
 $\lambda_{ii}=0$ and $\lambda_{12}=\lambda_{21}=-0.5$ resulting in the $s_\pm$ superconducting state\cite{ChubukovSpm,MazinSpm} .
 Such a model has been used to describe  
 an unconventional convex behaviour of $H_{c2}(T)$ observed experimentally in iron-based compounds.\cite{GurevichFeAsNature} 
 Here we suggest that an independent and more sensitive test for the multiband physics in iron pnictides can be implemented by 
 measuring temperature dependencies $\kappa_2(T)$. 
 As can be seen in Fig.(\ref{Fig:Spm}b)  
 $\kappa_2$ demonstrate a sharp increase at low temperatures even for not too small values of the ratio $D_1/D_2$
 and deviate strongly from the single-band behaviour shown by the green down-most curve.
 For the same parameters $H_{c2}(T)$ dependencies have only tiny deviations from the single-band one
 shown  by the green up-most line in  Fig.(\ref{Fig:Spm}a).
          
 Summarizing the above examples one can see that even a moderate disparity of diffusion constants when $D_1/D_2\sim 1$ 
 in two-band superconductors results in a significant increase of $\kappa_2(T)$ at low temperatures as compared to its value 
 at the critical temperature $\kappa_2(T_c)=\kappa_{GL}$. In result the magnetization slope $dM_z/dH$ becomes much less steep 
 as compared to single-band superconductors. This behaviour is illustrated in Fig.(\ref{Fig:Magnetization} ) 
 which shows the slopes $dM_z/dH$ normalized to their values at $T_c$ as functions of temperature for different values of $D_1/D_2$.  
 In both (a) and (b) panels the up-most green curve shows a single-band behaviour which is reproduced universally for 
 $D_1=D_2$ irrespective of the coupling parameters. The changes in magnetization slopes can be directly measures and yield an 
 important information about multiband pairing and diffusion constants in different bands.    
    
 \section{Discussion.} \label{Sec:Discussion} 
 The convex shape of $H_{c2}(T)$ curves is often considered as a signature of multiband pairing\cite{GurevichFeAsNature,BalatskyEdge,Hc2SrEu}. 
 However as demonstrated by the above two-band examples the conditions for having pronounced convexity such as shown in 
 Fig.(\ref{Fig:MgB2}c) are quite restrictive. If the disparity of diffusivities is not 
 extreme $D_1/D_2 \sim 1$ then deviations of $H_{c2}(T)$ from the conventional single-band theory 
 are not significant. However even in this case it is possible to 
 detect signatures of multiband pairing in the magnetic response 
 measuring the magnetization slope at high fields. 
 Shown in the right columns of Figs.(\ref{Fig:MgB2},\ref{Fig:Spm}) $\kappa_2(T)$ dependencies are quite sensitive to variations of
 diffusivities even in the range of parameters when $H_{c2}(T)$ curves look almost the same as the single-band one. 
  
  To understand qualitative features of $\kappa_2(T)$ in multiband superconductors it is instructive to consider several characteristic cases.
 First let us recover single-band results for $H_{c2}$ and $\kappa_2$ \cite{} assuming all diffusivities to be equal
  $D_k=D$ for $k=1...N$. 
 In this case $\rho_k=\rho$ so that  Eq.(\ref{Eq:Hc2}) reduces to $\hat A= \hat\Lambda^{-1}- \hat I [G_0 - U(\rho) -\ln (T/T_c) ]$. 
 The solvability condition ${\rm det}\hat A =0 $ yields a single-band equation for the upper critical 
 field\cite{SingleBandHc2Werthamer,SingleBandHc2Maki}  $U(\rho) +\ln (T/T_c) =0$. 
 The corresponding eigen vector is temperature independent and determined by the equation 
 $(\hat \Lambda ^{-1} - \hat I G_0) {\bm b} = 0$. 
 Then taking into account that $\kappa_2(T_c)=\kappa_{GL}$ from Eq.(\ref{Eq:kappa2Res}) we obtain the analytical expression 
 $\kappa_2^2/\kappa_{GL}^2= -\pi^4 \psi^{\prime\prime}/[56\zeta(3) \psi^{\prime 2}] $ 
 coinciding with the single-band result\cite{Caroli}. 
  
 To explain significant variations of $\kappa_2$ let us compare a 
 low- and high-temperature asymptotic of the Eq.(\ref{Eq:kappa2Res}):
 \begin{align} \label{Eq:kappa2Tc}
 \kappa_2 (T_c)= \sqrt{\frac{7\zeta(3)}{2\pi^5e^2} \frac{\sum_{k}\nu_k b_k^4}{(\sum_k \nu_kD_kb_k^2)^2 } }, \\ \label{Eq:kappa2T0}
 \kappa_2 (0) = \sqrt{ \frac{1}{32\pi e^2} \frac{\sum_{k}\nu_kb^4_k D_k^{-2}}{(\sum_k \nu_kb_k^2)^2} },
 \end{align}             
 where we have used that $\psi^{\prime}_k \approx \rho^{-1}_k$, $\psi^{\prime\prime}_k \approx -\rho^{-2}_k$ at $T\to 0$ and $\psi^\prime_k= \pi^2/2$, 
 $\psi^{\prime\prime}_k =- 14\zeta(3)$ at $T=T_c$.  

 Eqs.(\ref{Eq:kappa2Tc},\ref{Eq:kappa2T0}) demonstrate that in the limit of a strong disparity between diffusivities  the value of 
 $\kappa_2 (T_c)$ is determined by the maximal diffusivity 
 while $\kappa_2 (0)$ is determined by the minimal one.
 Therefore $\kappa_{GL}=\kappa_2(T_c)\sim 1/D_1$
 and $\kappa_2(0)\sim 1/D_2$ so that the low temperature increase of $\kappa_2(0)/\kappa_{GL} \sim D_1/D_2 \gg 1$ is 
 determined by the ratio
 of maximal and minimal diffusivities $D_1=\max(D_k)$ and $D_2=\min(D_k)$ respectively. 
   
 Such a behaviour can be qualitatively understood as follows. Near the critical temperature 
 Eq. (\ref{Eq:Hc2}) reduces to $\hat A= \hat \Lambda^{-1} - \hat I G_0$ so that gap amplitudes $b_k$ are determined 
only by the coupling matrix.  The magnetic field is small so that $\rho_k\ll 1$ and its influence on the gap amplitudes is negligible. 
 However the contributions to superconducting current and magnetization (\ref{Eq:Current},\ref{Eq:MagnetizationMultiband}) from each band
 are proportional to the corresponding diffusion coefficients. Hence in the limit of strong disparity $D_1/D_2 \ll (\gg ) 1$ the 
 band with the largest diffusivity provides a dominant contribution to the magnetization near $T_c$. 
 On the other hand at low temperatures the magnetic field is large so that $\rho_k\gg 1$ and therefore
 can effectively suppress superconducting correlations.
 From Eq.(\ref{Eq:Coeffitients}) one can see that 
 the anomalous function amplitude
 is smaller in bands with larger diffusivities.
 Hence at $T\to 0$ the most significant contribution to the magnetic response and $\kappa_2$ is determined by the band with the smallest diffusivity.    
 
 Finally, the calculations presented in this paper consider only the orbital depairing mechanism and neglect paramagnetic
 effects which are important in iron pnictide compounds with large critical fields\cite{Paramagnetic-1,Paramagnetic-2,Paramagnetic-3,Paramagnetic-4,Paramagnetic-5}. 
 High values of $H_{c2}=\phi_0/(2\pi\xi^2)$ in these materials are determined by short coherence lengths $\xi\sim 1-3$ nm \cite{GurevichFFLO}   
 which are not consistent with the dirty limit approximation considered here. It is possible however to develop 
 a theory for $\kappa_2$ in superconductors with arbitrary impurity concentration\cite{EilenbergerKappa2, KitaKappa2}. 
 In single-band superconductors $\kappa_2(T\to 0)$ diverges in the clean limit but for experimentally relevant finite impurity concentrations 
 the changes are not dramatic as compared to the dirty limit\cite{Kappa2Exp1}. 
 On the other hand in the multiband case an interplay between different Fermi velocities and impurity scattering rates 
 should result in a non-trivial modifications of $\kappa_2(T)$ temperature dependencies.   
          
 \section{Conclusion}   
 \label{Sec:Conclusion}
 To conclude we have calculated the parameter $\kappa_{2}$ characterizing magnetization slopes $dM_z/dH$ 
 in dirty multiband superconductors at high fields $H_{c2}-H \ll H_{c2}$. 
 The developed theory describes any number of superconducting bands and
 arbitrary set of pairing constants. We have shown quite generally that in contrast to the dirty single-band superconductors 
 the temperature dependencies of $\kappa_2(T)$ have remarkable features which are highly sensitive to the 
 multiband effects.
 The low-temperature increase of $\kappa_2$ as compared 
 to its value at $T_c$ is found to be strongly pronounced even for the moderate disparity of diffusion coefficients in different bands. 
 This effect should be particularity appealing for experimental identification since it could unambiguously  
 confirm unconventional magnetic behaviour of multiband superconductors. 
 We have considered several examples of two-band materials like MgB$_2$ and iron pnictides and demonstrated that 
 $\kappa_2$ is much more sensitive than $H_{c2}$ to the ratio of diffusion coefficients in different bands.
 The established relations between $\kappa_2$ and gap function amplitudes provide a basis to study 
 thermodynamic and transport  properties of multiband superconductors in high magnetic fields.     
 
 \section{Acknowledgements}
  The author is grateful to Egor Babaev and Alex Gurevich for the useful discussions. 
  The work was supported by the Swedish Research Council Grants
No. 642-2013-7837.
         
 \appendix

  \section{Proof of the relation Eq.(\ref{Eq:Nontrivial})} \label{Eq:Appendix}
  We use the relations $\nabla= (\bm \partial_+ + \bm \partial_-)/2 $ introducing the operators 
  $\partial_\pm = {\bm x} \partial_x + {\bm y}(\partial_y \pm 2ie A_y )$ so that
  $\partial_\pm^2 = \partial_x^2 + (\partial_y \pm 2ie A_y ) ^2 $.
  The gap functions satisfy 
  \begin{eqnarray}
  \partial_+^2 \Delta^* = -2eH_{c2}\Delta^*   \\
  \partial_-^2 \Delta = -2eH_{c2}\Delta
  \end{eqnarray}   
  Due to the relations
  \begin{eqnarray}
  &\partial_x \Delta = i (\partial_y - 2ie A_y ) \Delta \\
  &\partial_x \Delta^* = - i (\partial_y + 2ie A_y )\Delta^*
 \end{eqnarray}  
  we get 
 \begin{equation} \label{Eq:ZeroSquare}
 (\partial_- \Delta)^2=0; \;\; (\partial_+ \Delta^*)^2=0
 \end{equation}  
 Then the average is given by 
 $$
 4\langle |\Delta|^2 \nabla^2 |\Delta|^2 \rangle = 
 \langle |\Delta|^2 (\partial_+^2 + \partial_-^2 + 2 \partial_+\partial_- ) |\Delta|^2 \rangle .
 $$
 Let us consider the three terms in separate
 
 \begin{align} {\bf (i)}
 &\langle |\Delta|^2 \partial_+^2 |\Delta|^2\rangle  = \\ \nonumber
 &\langle |\Delta|^2 \left( \Delta^*\partial_+^2 \Delta + \Delta\partial_+^2 \Delta^* + 2 \partial_+ \Delta \partial_+ \Delta^*\right) \rangle = 
 \\ \nonumber
 & -2eH_{c2}  \langle |\Delta|^4 \rangle + 
  \langle |\Delta|^2\Delta^*\partial_+^2 \Delta \rangle + 
  2 \langle |\Delta|^2 \partial_+ \Delta \partial_+ \Delta^* \rangle  
 \end{align}
  \begin{align} {\bf (ii)}
 &\langle |\Delta|^2 \partial_-^2 |\Delta|^2\rangle  = \\ \nonumber
 &\langle |\Delta|^2 \left( \Delta^*\partial_-^2 \Delta + \Delta\partial_-^2 \Delta^* + 2 \partial_- \Delta \partial_- \Delta^*\right) \rangle = 
 \\ \nonumber
 & - 2eH_{c2}  \langle |\Delta|^4 \rangle + 
 \langle |\Delta|^2 \Delta\partial_-^2 \Delta^* \rangle + 
 2 \langle |\Delta|^2\partial_- \Delta \partial_-\Delta^*  \rangle  
 \end{align}
  \begin{align} {\bf (iii)} 
 & -2\langle |\Delta|^2 \partial_+\partial_- |\Delta|^2\rangle  = \\ \nonumber
 & \langle (\partial_+ |\Delta|^2 )^2 \rangle\rangle + \langle (\partial_- |\Delta|^2 )^2 \rangle = \\ \nonumber
 & \langle (\partial_+ \Delta )^2 \Delta^{*2} \rangle + \langle (\partial_+ \Delta^*)^2 \Delta^{2} \rangle + 
 2 \langle |\Delta|^2 (\partial_+\Delta)(\partial_+\Delta^*)\rangle + \\ \nonumber
 &\langle (\partial_- \Delta )^2 \Delta^{*2} \rangle + \langle (\partial_- \Delta^*)^2 \Delta^{2} \rangle + 
 2 \langle|\Delta|^2 (\partial_-\Delta)(\partial_-\Delta^*) \rangle= \\ \nonumber
 & \langle (\partial_+ \Delta )^2 \Delta^{*2} \rangle + 
 2 \langle |\Delta|^2 (\partial_+\Delta)(\partial_+\Delta^*) \rangle + \\ \nonumber
 & \langle (\partial_- \Delta^*)^2 \Delta^{2} \rangle +
  2  \langle|\Delta|^2 (\partial_-\Delta)(\partial_-\Delta^*) \rangle
 \end{align}
 where we took into account the Eqs.(\ref{Eq:ZeroSquare}).
 Collecting all terms we get 
  \begin{align} 
 &4\langle |\Delta|^2 \nabla^2 |\Delta|^2 \rangle =  -4eH_{c2}  \langle |\Delta|^4 \rangle + \\ \nonumber
 &\langle \Delta^{*2} [ \Delta \partial_+^2 \Delta - (\partial_+ \Delta)^2] \rangle + 
  \langle\Delta^{2} [ \Delta^* \partial_-^2 \Delta^* - (\partial_- \Delta^*)^2] \rangle  
  \end{align}
  The last two terms here can be transformed in a similar way as follows 
  \begin{align}
  &\Delta \partial_+^2 \Delta - (\partial_+ \Delta)^2  = \\ \nonumber
  &\Delta \partial_-^2 \Delta - (\partial_- \Delta)^2 = -4eH_{c2} \Delta^2 \\
  &\Delta^* \partial_-^2 \Delta^* - (\partial_- \Delta^*)^2  = \\ \nonumber
  &\Delta^* \partial_+^2 \Delta^* - (\partial_+ \Delta^*)^2 =  -4eH_{c2} \Delta^{*2}   
  \end{align}
  so that finally  we get  
  $$
  \langle |\Delta|^2 \nabla^2 |\Delta|^2 \rangle =  -2eH_{c2}  \langle |\Delta|^4 \rangle
  $$
  which proves the Eq.(\ref{Eq:Nontrivial}).


\begin{thebibliography}{99} 
  
 \bibitem{MgB2:Discovery} 
 J. Nagaamatsu, N. Nakagawa, T. Murakana, Y. Zenitani, J. Akimitsu, Nature {\bf 410}, 63 (2001).
   
 \bibitem{MgB2:MazinPRL} 
 J. Kortus, I. I. Mazin, K. D. Belashchenko, V. P. Antropov, and L. L. Boyer, Phys. Rev. Lett. {\bf 86}, 4656 (2001).    
  
 \bibitem{MgB2:Mazin} 
 I. I. Mazin and V. P. Antropov, Physica (Amsterdam) {\bf  385C}, 49 (2003).
 
 \bibitem{Sr2RuO4} 
 A. Damascelli, D. H. Lu, K. M. Shen, N. P. Armitage,  
 F. Ronning, D. L. Feng, C. Kim, Z.-X. Shen, T. Kimura, Y. Tokura, 
 Z. Q. Mao, and Y. Maeno, Phys. Rev. Lett. {\bf 85}, 5194 (2000).

 \bibitem{FeAsExp1} 
 Y. Kamihara, T. Watanabe, M. Hirano, and H. Hosono, J. Am. Chem. Soc. {\bf 130}, 3296 (2008).
 
  \bibitem{MazinSpm} 
 I. I. Mazin, D. J. Singh, M. D. Johannes, and M. H. Du, Phys. Rev. Lett. {\bf 101}, 057003 (2008).
 
 \bibitem{FeAsMultiband}
 K. Kuroki, S. Onari, R. Arita, H. Usui, Y. Tanaka, H. Kontani, and H. Aoki, Phys. Rev. Lett. {\bf 101}, 087004 (2008).

 \bibitem{ChubukovSpm}
 A. V. Chubukov, D. V. Efremov, and I. Eremin, Phys. Rev. B {\bf 78}, 134512 (2008).
 
  \bibitem{MgB2:abinitio}
 A.Y. Liu, I.I. Mazin, and J. Kortus, Phys. Rev. Lett. {\bf 87}, 087005 (2001);
 H.J. Choi, D. Roundy, H. Sun, M.L. Cohen and S.G. Louie, Nature {\bf 418}, 758 (2002).  
 
 \bibitem{MgB2:TopicalReview}
 C. Buzea and T. Yamashita, Supercond. Sci. Technol. {\bf 14}, R115 (2001).
 
 \bibitem{FeSeThinFilms}
 Jian-Feng Ge et al., Nature Mat. {\bf 14}, 285 (2015).
 
 \bibitem{Moshchalkov15Main}
 V. Moshchalkov, M. Menghini, T. Nishio, Q. H. Chen, A. V. Silhanek, V. H. Dao, 
 L. F. Chibotaru, N. D. Zhigadlo,  and J. Karpinski Phys. Rev. Lett. {\bf 102}, 117001 (2009). 
 
 \bibitem{Moshchalkov15}
 J. Gutierrez, B. Raes, A. V. Silhanek, L. J. Li, N. D. Zhigadlo, J. Karpinski, J. Tempere, and V. V. Moshchalkov
 Phys. Rev. B {\bf 85}, 094511 (2012);
 Taichiro Nishio, Vu Hung Dao, Qinghua Chen, Liviu F. Chibotaru, Kazuo Kadowaki, and Victor V. Moshchalkov
 Phys. Rev. B {\bf 81}, 020506(R) (2010); 

 \bibitem{Bending15}
 S. J. Ray, A. S. Gibbs, S. J. Bending, P. J. Curran, E. Babaev, C. Baines, A. P. Mackenzie, and S. L. Lee
 Phys. Rev. B {\bf 89}, 094504 (2014).
 
 \bibitem{BabaevSpeight}
 E. Babaev and M. Speight, Phys. Rev. B {\bf 72}, 180502R (2005).
 
 \bibitem{SilaevBabaevPRB2011} M. Silaev, E. Babaev, Phys. Rev. B {\bf 84}, 094515 (2011).
 
 \bibitem{CarlstromPRL2010} E. Babaev, J. Carlstrom, M. Speight, Phys. Rev. Lett. {\bf 105}, 067003 (2010). 
 
   \bibitem{GurevichFeAsNature}
 F. Hunte, J. Jaroszynski, A. Gurevich, D. C. Larbalestier, R. Jin, 
 A. S. Sefat, M. A. McGuire, B. C. Sales, D. K. Christen, D. Mandrus,
 Nature {\bf 453}, 903 (2008).
 
 \bibitem{Exp60T-1} 
 J. Jaroszynski et al., Phys. Rev. B {\bf 78}, 174523 (2008).

 \bibitem{Exp60T-2}
 A. Yamamoto et al., Appl. Phys. Lett. {\bf 94}, 062511 (2009).
 
 \bibitem{Exp60T-3}
 G. Fuchs et al., Phys. Rev. Lett. {\bf 101}, 237003 (2008).
 
 \bibitem{Exp60T-4}
 M. M. Altarawneh et al., Phys. Rev. B {\bf 78}, 220505 (2008).

 \bibitem{Exp60T-5}
 M. Fang  et al., Phys. Rev. B {\bf 81}, 020509 (2010).
 
   \bibitem{Paramagnetic-1}
 S. Khim, et al., Phys. Rev. B {\bf 81}, 184511 (2010).
 
 \bibitem{Paramagnetic-2}
 H. Lei, et al., Phys. Rev. B {\bf 81}, 184522 (2010).

 \bibitem{Paramagnetic-3}
 M. Kano, et al., J. Phys. Soc. Jpn. {\bf 78}, 084719 (2009).

 \bibitem{Paramagnetic-4}
 T. Terashima, et al., J.Phys. Soc. Jpn. {\bf 78}, 063702 (2009).
 
 \bibitem{Paramagnetic-5}
 D. Braithwaite, G. Lapertot, W. Knapo, and I. Sheikin, J. Phys.
 Soc. Jpn. {\bf 79}, 053703 (2010). 
 
  \bibitem{GurevichFFLO}
 A. Gurevich, Phys. Rev. B {\bf 82} 184504 (2010).
 
  \bibitem{MgB2Ex1} V. Braccini et al., Phys. Rev. B {\bf 71}, 012504 (2005).  
  
   \bibitem{GurevichMgB2}
 A. Gurevich, Phys. Rev. B {\bf 67}, 184515 (2003); 
 
 \bibitem{GurevichPhysicaC}
 A. Gurevich, Physica C {\bf 456}, 160-169 (2007);
 
  \bibitem{KoshelevGolubovHc2}
 A. A. Golubov and A. E. Koshelev, Phys. Rev. B {\bf 68}, 104503 (2003).
 
  \bibitem{SingleBandHc2Maki}
 K. Maki, Phys. Rev. {\bf 148}, 362 (1966).
 
  \bibitem{SingleBandHc2Werthamer}
 N.R. Werthamer, E. Helfand, and P.C. Hohenberg, Phys. Rev. {\bf 147}, 288 (1966).
 
  \bibitem{MgB2Ex2} V. Ferrando et al., Phys. Rev. B {\bf 68} 094517 (2003).
 \bibitem{MgB2Ex3} F. Bouquet et al., Physica C {\bf 385} 192 (2003).
 \bibitem{MgB2Ex4} A. Gurevich et al., Supercond. Sci. Technol. {\bf 17}, 278 (2004).
 
  \bibitem{Hc2SrEu} R. Hu, E. D. Mun, M. M. Altarawneh, C. H. Mielke, V. S. Zapf, S. L. Bud'ko, and P. C. Canfield, 
  Phys. Rev. B {\bf 85}, 064511 (2012).

 \bibitem{Hc2CaLaFeAs} 
 W Zhou, et al., Supercond. Sci. Technol. {\bf 26} 095003 (2013).  
 
    \bibitem{BalatskyEdge}   
 J. M. Edge, A. V. Balatsky, J. of Supercond. and Nov. Magn., {\bf 28}, 2373 (2015).
 
   \bibitem{concave-1}
S. I. Vedeneev et al., Phys. Rev. B {\bf 87}, 134512 (2013).

 \bibitem{Maki}
 K. Maki. Physics, {\bf 1} 21 (1964).
 
  \bibitem{AbrikosovParameter}
 W.H. Kleiner, L.M. Roth, S.H. Autler, Phys. Rev. {\bf A133} 1226 (1964).

  \bibitem{MakiTsuzukiKappa2}
 K. Maki and T. Tsuzuki, Phys. Rev. {\bf 139} A868 (1965).
 
 \bibitem{Caroli}
 C. Caroli, M. Cyrot, P.G. de Gennes, Solid State Comm. {\bf 4} 17 (1966).
 
 \bibitem{UsadelKappa2} Klaus. D. Usadel, Phys. Rev. B {\bf 4}, 99 (1971).
 
  \bibitem{EilenbergerKappa2}
 G. Eilenberger, Phys. Rev. {\bf 153} 584 (1967).
 
  \bibitem{KitaKappa2}
 T. Kita, Phys. Rev. B {\bf 68}, 184503 (2003).
 
  \bibitem{UsadelKappa2Strong} Klaus. D. Usadel, Phys. Rev. B {\bf 2}, 135 (1970).
  
   \bibitem{Kappa2Exp1} C.W. Smith, E.R. Sanford and R.W. Rollins, Phys. Lett. {\bf 27A}, 362 (1968).
 \bibitem{Kappa2Exp2} E. Fischer and H.P. Vieli, Phys. Lett. {\bf 26A}, 35 (1967).
 \bibitem{Kappa2Exp3} W.A. Fietz and W.W. Webb, Phys. Rev. {\bf 161}, 423 (1967).
 
  \bibitem{MakiSpecificHeat}
 K. Maki, Phys. Rev., {\bf 139}, A702 (1965). 
 
  \bibitem{CaroliMakiFluxFlow}
 K. Maki, Phys. Rev. {\bf 141}, 331 (1966). 
   
  \bibitem{ThompsonFluxFlow}
 R.S. Thompson, Phys. Rev. B, {\bf 1}, 327 (1969).
 
 \bibitem{Schon}
 G. Sch\"{o}n in {\it Modern Problems in Condensed Matter Sciences: Nonequilibrium Superconductivity}
 eds. D.N. Langenberg and A.I. Larkin, p. 589-640, Elsevier (1986).
 
   \bibitem{BrandtOvchinnikov}
  Yu.N. Ovchinnikov and E.H. Brandt, Phys. Stat. Sol.(b) {\bf 67}, 301 (1975) .
  
   \bibitem{MgB2:constants} 
 A.A. Golubov, J. Kortus, O.V. Dolgov, O. Jepsen, Y. Kong, O.K. Andersen, B.J. Gibson, K. Ahn, and R.K. Kremer,
 J. Phys.: Condens. Matter {\bf 14}, 1353 (2002).
  
 
 
%
%
%
%
%
%
%
%
     
         
 \end{thebibliography}
 \end{document}